\newcommand{\citet}[1]{\cite{#1}}

\documentclass{article}

\date{}

\usepackage[caption=false]{subfig}
\usepackage{arxiv}

\usepackage{microtype}
\usepackage{booktabs}

\usepackage{amsmath}
\usepackage{float}
\usepackage{xcolor}
\usepackage[hidelinks]{hyperref}

\usepackage{graphicx}
\usepackage{authblk}

\renewcommand{\undertitle}{This paper has been accepted for publication at\\ DeceptAI: International Workshop on Deceptive AI}

\begin{document}

\title{Learning to Deceive in Multi-Agent Hidden Role Games}
%
%
\author[1]{Matthew Aitchison}
\author[2]{Lyndon Benke}
\author[1]{Penny Sweetser}


\affil[1]{
    ANU College of Engineering and Computer Science\\
    The Australian National University
    }
\affil[2]{
    Defence Science and Technology Group\\
    Department of Defence
}
%

%

%
\maketitle              
\begin{abstract}
Deception is prevalent in human social settings. However, studies into the effect of deception on reinforcement learning algorithms have been limited to simplistic settings, restricting their applicability to complex real-world problems. This paper addresses this by introducing a new mixed competitive-cooperative multi-agent reinforcement learning (MARL) environment, inspired by popular role-based deception games such as Werewolf, Avalon, and Among Us. The environment's unique challenge lies in the necessity to cooperate with other agents despite not knowing if they are friend or foe. Furthermore, we introduce a model of deception which we call \textit{Bayesian belief manipulation} (BBM) and demonstrate its effectiveness at deceiving other agents in this environment, while also increasing the deceiving agent's performance.

\end{abstract}

\section{Introduction}


Human tendency to trust in the honesty of others is essential to effective cooperation and coordination \cite{levine2014truth}. At the same time, trust opens up the risk of being exploited by a deceptive party. Deception is prevalent in many real-life settings, especially in those that are adversarial in nature. Examples include cyber-security attacks \cite{almeshekah2016cyber}, warfare \cite{daniel2013strategic}, games such as poker \cite{brown2019deep}, and even every-day economic interactions \cite{gneezy2005deception}. Outside of simplistic signaling games (e.g. \citet{la2016deceptive,carroll2011game}), use, and defence against, deceptive strategies has received relatively little research attention. 

The popularity of social deception games, such as Werewolf, Avalon, and Among Us, reveals both a human fascination with deception, and the challenges that it creates. These games require players to cooperate with their \textit{unknown} teammates to achieve their goals, all while trying to hide their identities. Therefore, social deception games could provide an interesting setting for research into deception. 
To explore deception beyond simple signalling games, we introduce a new hidden role mixed competitive-cooperative multi-agent reinforcement learning (MARL) environment \textit{Rescue the General} (RTG). Our environment features three teams: red and blue, who have conflicting goals, and a neutral green team. To perform well, agents must learn subtle trade-offs between acting according to their team's objectives and revealing too much information about their roles. 

To this end, we trained two sets of agents to play RTG, one in which the agent pursues the rewards of the game as per normal, and another where the agents are provided an intrinsic reward \cite{schmidhuber2010formal} to incentivise deceptive actions. We compare the behaviour of the two groups and analyse the their performance when pitted against each other.\footnote{Source code for the environment and model, including a script to reproduce the results in this paper, can be found at \url{https://github.com/maitchison/RTG}.} The main contributions of this work are:


\begin{enumerate}
    \item The introduction of a new deception-focused hidden-role MARL environment.
    \item A new model for deception called \textit{Bayesian belief manipulation} (BBM).
    \item Empirical results showing the effect of BBM on agents' behaviour, and which demonstrate its effectiveness as a deception strategy.
\end{enumerate}

\section{Background and Related Work}


Previous game-theoretic approaches have typically modelled deception through signalling \cite{ho1978teams}, where one player can, at a cost, send a signal conveying false information. An example in network security, examined by \citet{carroll2011game}, has a defender who may attempt to deceive an attacker by disguising honeypots as regular computers.\footnotemark{} Other work has explored the evolution of deceptive signalling in competitive-cooperative learning environments, \citet{floreano2007evolutionary} found that teams of robots in competitive food-gathering experiments spontaneously evolved deceptive communication strategies, reducing competition for food sources by causing harm to opponents. 
Our research differs from these signal based deception studies in that instead of modelling deception as an explicit binary action, we require agents to learn complex sequences of actions to mislead, or partially mislead, other players.

\footnotetext{A honeypot is a networked machine designed to lure attackers, but which contains no real valuable information.}


An extension to game theory is hypergame theory \cite{bennett1980hypergames}. Hypergame theory models games where players may be uncertain about others players' preferences (strategies), and therefore may disagree on what game they are playing. Because the model includes differences in agents' perception of the game, hypergame theory provides a basis for modelling misperception, false belief, and deception \cite{kovach2015hypergame}. Examples of hypergame analysis in practice include \citet{vane2002using}, who consider deception in single-stage, normal form hypergames, and \citet{gharesifard2013stealthy} who model deception about player preferences for games in which the deceiver has complete knowledge of the target. Hypergames have also been used to model the interaction between attackers and defenders in network security problems \citet{gutierrez2015modeling}. Our environment can be seen as a hypergame because players do not know the roles of other players in the game. However, unlike previous work, where all players' actions were fully observable, our environment includes partially observed actions increasing the potential for disagreement and deception.

The most similar work to our own is \citet{strouse2018learning} who also incentivise agents to manage information about their roles. They use mutual information between goal and state as a regularisation term during optimisation to encourage or discourage agents from revealing their goals. Our work differs in that we instead incorporate deception into the agents' rewards, which allows agents to factor the future possibility of deception into their decision-making process. 

Our deception model is most similar to the Bayesian model proposed by \citet{ettinger2010theory}, who consider agents with incomplete information about the types of other players. Unlike our work, this approach assumes a two-player game with fully observable actions, in which the history of each player is known to both players. Our approach enables the modelling of deception in multi-agent games with partial observability, in which player histories are unknown and must be estimated from local observations.

Reasoning about other players' roles under uncertainty has also been explored by \citet{serrino2019finding}, who use counterfactual regret minimisation to deductively reason about other players' roles in the game of Avalon. Their work is similar in that their agents must identify unknown teammates' roles. However, their approach differs in that they do not explicitly encourage deception.
\citet{macnally2018action} consider the broader problem of communicating intent in the absence of explicit signalling. An online planner is used to select actions that implicitly communicate agent intent to an observer. This approach has been applied to deception by \citet{masters2017deceptive}, by maximising rather than minimising the difference between agent and observer beliefs. Unlike our work, these approaches assume full observability, and require a model of the environment for forward planning.

While many existing MARL environments have explored cooperation between teammates (\cite{foerster2016learning,samvelyan2019starcraft,leibo2017multi,iqbal2019actor,baker2019emergent} only ProAvalon \cite{serrino2019finding} requires cooperation with \textit{unknown} teammates. However, ProAvalon is not suitable for our needs, due to the high degree of shared information.\footnote{In Avalon only player roles, and who decided to sabotage a mission are hidden.} In contrast, our environment, RTG, with limited player vision allows agents to have very different belief about game's current state. This difference is important for our BBM model as modelling other players (potentially incorrect) understanding of the game's current state is essential to manipulating their belief. 

\section{Rescue the General}

We developed a new MARL environment, RTG, which requires agents to learn to cooperate and compete when teammates are unknown. Popular hidden role deception games inspired many of the environment's core mechanics. RTG is open source, written in Python and uses the Gym framework \cite{brockman2016openai}. 

We designed RTG with the following objectives: the game should be fast to simulate but complex to solve; game observations should be both human and machine-readable; code for the environment should be open source and easy to modify; the game should be mixed competitive/cooperative (i.e., not zero-sum); hidden roles should be a core game mechanic;\footnote{In this paper, role refers to the policy a player follows and maps directly to the player's team. Different roles may exist within a single team in other games.} and good strategies should require non-trivial (temporally extended) deception.

\subsection{Teams and Objectives}

The game consists of three teams: red, green, and blue. The blue team knows the location of a general and must perform a rescue by dragging them to the edge of the map, but requires multiple soldiers to do so. Green are `lumberjacks' who receive points for harvesting wood from the trees scattered around the map, and whose interests are orthogonal to the other teams. The red team does not know the general's location and must find and kill the general. Similar to the StarCraft Multi-Agent Challenge \cite{samvelyan2019starcraft}, each soldier is controlled independently, and has a limited vision (a range of six tiles for red, and five for green and blue). The complication is that no soldier knows any other's identity', including their own team members. Communication is limited to the soldiers' actions, namely movement by one tile north, south, east or west and the ability to shoot in one of the four cardinal directions. All soldiers have an ID colour allowing soldiers to track previous behaviour.

Player teams are randomised at the beginning of each game. Their locations are randomily initialised such that they are near to each other, but always more than 2-tiles away from the general. 

\subsection{Observational Space}

\begin{figure}[h]
    \center 
    \includegraphics[width=.85\textwidth]{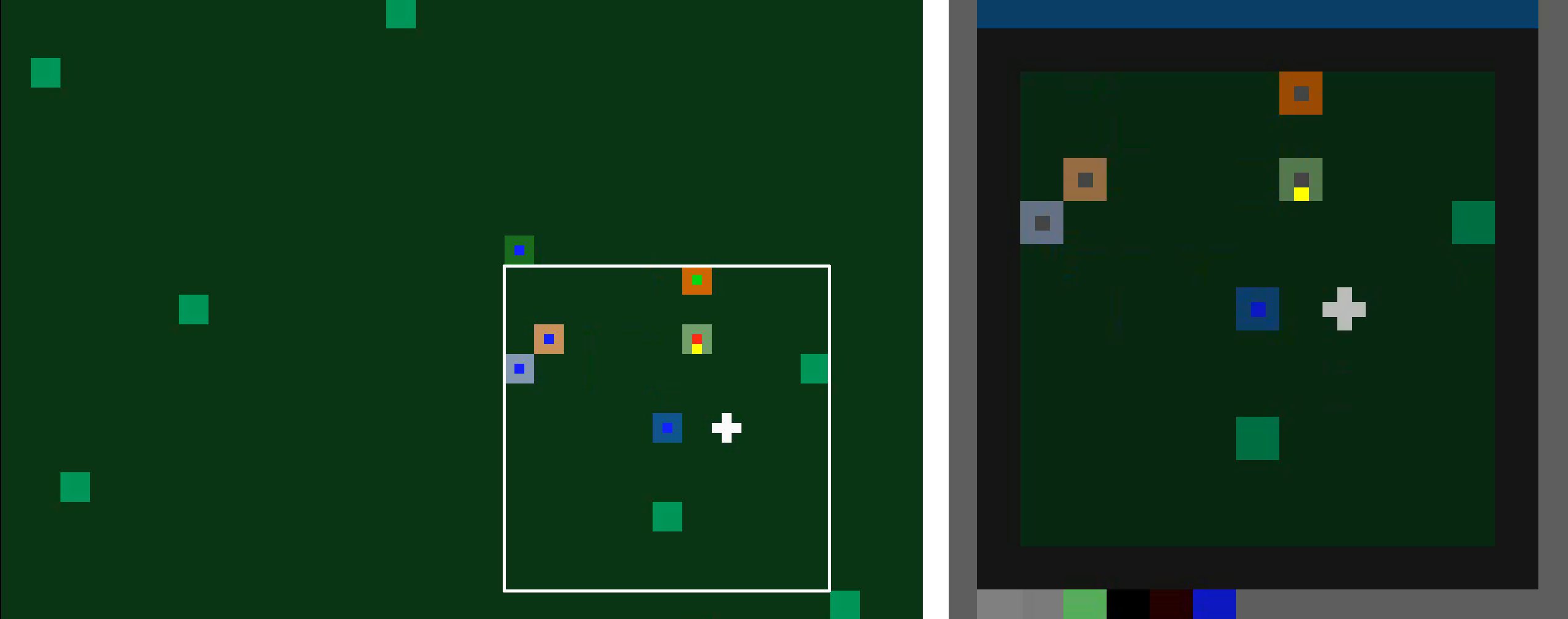} 
    \caption{The global observation (left) and an agent's local observation (right). Soldiers are painted with their unique ID color on the outside, and their team color on the inside. Local observations omit the team color for all non-red players. Each map tile is 3x3 and encoded in RGB colour.} 
    \label{fig:local_obs}
\end{figure}

Egocentric observations are provided to each agent in RGB format, as shown in Figure \ref{fig:local_obs}. Status indicator lights give the agent information on their x-location, y-location, health, turns until they can shoot, turns until a game timeout, and team colour. A marker indicating the direction of the general is also given to blue players. Rewards depend on the scenario and are detailed in Section \ref{sec:reward}. 

\subsection{Scenarios}

The RTG environment provides six scenarios of differing levels of challenge. These scenarios are given in Table \ref{tab:rtg}. Additional custom scenarios can also be created through configuration options (for details see Appendix A).

In the primary scenario, Rescue, the blue team must rescue a general without allowing a lone red player to find and shoot the general. The Wolf scenario features a single red player who must kill all three green players before being discovered. R2G2 demonstrates orthogonal objectives, with two green players harvesting trees while two red players attempt to find and kill the general. Red2, Green2 and Blue2 are simple test environments where each team can learn their objective unobstructed.

\begin{table}[h]
    \caption{Summary of scenarios provided the Rescue the General environment. Player counts are listed in red, green, and blue order.}
    \vskip 0.15in
    \centering
    \small
    \begin{tabular}{l c c c c}
         \hline
         Scenario & Players & Type & Roles & Challenge \\
         \hline
         Rescue &  1-1-4 & Mixed & Hidden & Hard \\
         Wolf & 1-3-0 & Competitive & Hidden & Moderate \\
         R2G2 & 2-2-0 & Mixed & Visible & Moderate \\
         Red2 & 2-0-0 & Cooperative & Visible & Easy \\
         Green2 & 0-2-0 & Cooperative & Visible & Easy \\
         Blue2 & 0-0-2 & Cooperative & Visible & Easy \\
         \hline
    \end{tabular}
    
    \label{tab:rtg}
\end{table}

\subsection{Reward Structure}
\label{sec:reward}

The rewards are structured such that a victory for red or blue always results in a score of ten, and are outlined in Table~\ref{tab:rewards}. $\text{R}_{\text{b}}$ refers to the total reward for the blue team this game, and is used to make sure blue always receives a maximum of ten points for a win. Green is also able to score ten points if they harvest all the trees on the map, but unlike red and blue, they will not terminate the game by achieving their objective. Due to the complexity of rescuing the general (two players must work together to move the general tile-by-tile) we provide small rewards to blue for completing partial objectives. 

\begin{table}[h]
    \centering
    \caption{Rewards for each team in the RTG game.}
    \vspace{0.5em}
    \small
    \begin{tabular}{p{5.0cm} c c c}
    \toprule
        Event & Red & Green & Blue\\
    \midrule
        General killed & 10 & 0 & -10\\
        General rescued & -10 & 0 & $10-\text{R}_{\text{b}}$ \\
        Timeout & 5 & 0 & -5 \\
        Green harvests tree & 0 & 1 & 0 \\
        Blue next to general (first time) & 0 & 0 & 1 \\
        General moved closer to map edge & 0 & 0 &  $\frac{(10-\text{R}_{\text{b}})}{20}$ \\
    \bottomrule
    \end{tabular}
    \label{tab:rewards}
\end{table}

\section{Bayesian Belief Manipulation Model}

Our deception model is based on belief manipulation \cite{eger2017practical}. Each agent keeps track of what it believes other agents believe that an action would imply about their role. We refer to this as `inverse' prediction. That is, predicting what others would predict about ourselves. Using these inverse predictions and an assumption of Bayesian belief updates, a reward is generated to incentivise agents to take actions that would mislead other agents about their true role. We call this Bayesian belief manipulation (BBM).

\subsection{Theory of Mind} 

\begin{figure} 
    \centering 
    \includegraphics[width=0.25\textwidth]{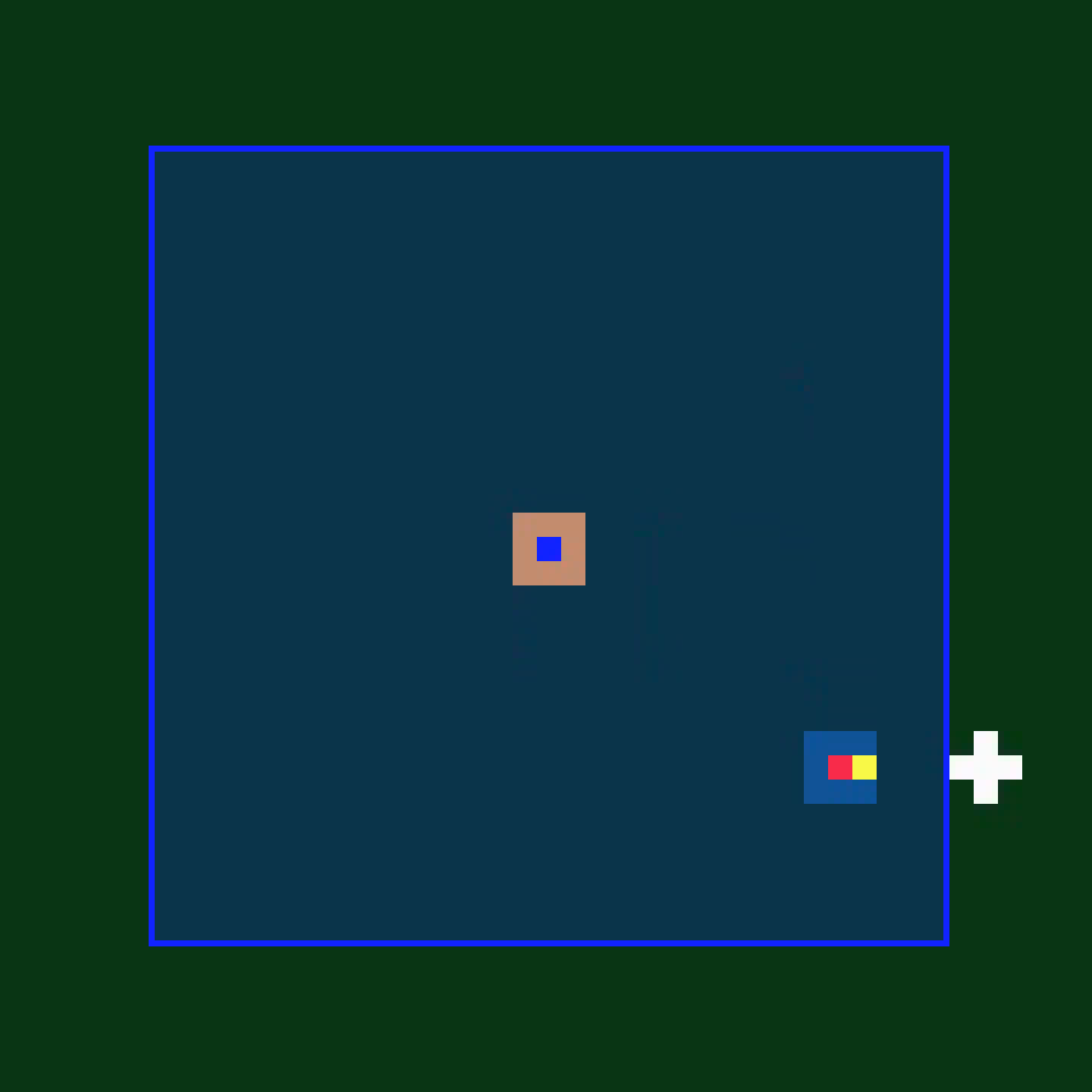} 
    \caption{Should red believe they are giving away their role by shooting east? This depends on whether red knows that blue knows that red is shooting the general (depicted with a cross). Blue's limited vision is shaded, and both soldier's roles are visualised for context. If blue knows that red knows where the general is, blue will interpret this as a very `red' move.} 
    \label{fig:theory_of_mind} 

\end{figure}

In order to deceive another player, we need to model our belief about their belief of how our actions would be interpreted. Take, for example, the situation in Figure \ref{fig:theory_of_mind}. If red believes that blue knows that red can see the general, this action will be considered a very `red' move. However, if red believes that blue is unaware of the general, this action can, perhaps, be taken without revealing red's role. Thus, agents must model, not only the beliefs about others but also others' beliefs about themselves. 

\subsection{Assumptions}
\label{sec:assumptions}
In our model the deceiving agents operate under the following assumptions:

\vfill

\begin{enumerate}
    \item All non-deceiving agents are Bayesian in their approach to updating belief about roles.\footnote{Our agents do not explicitly use Bayesian updates when estimating the roles of other players, but we find this assumption to be an effective model.}
    \item The policy of each role is known by all players.\footnote{We partially relax this assumption, see Section \ref{sec:dap}.}
    \item Agents are limited to second-order theory of mind.
\end{enumerate}

The second assumption amounts to knowing the type of the policy, that is, knowing the set of potential policies, but not knowing which one the agent is acting out. Because the set of policies could, in theory, be very large, this assumption is not as limiting as it sounds. 
The third assumption follows the findings of \citet{de2013much}, who conclude that while agents with first-order and second-order theory of mind (ToM) outperform opponents with shallower ToM, deeper levels of recursion show limited benefit.

\subsection{Notation}

We use the following notation:

\begin{itemize}
    \item $A_i$ refers to the $i^{th}$ agent, and $A_i^{z}$ refers to the event that the $i^{th}$ agent has role $z$.
    \item $h_i^t$ is $A_i$'s history of local observations up to and including timestep $t$. Where $t$ is omitted it can be assumed to be the current timestep.
    \item $h_{i,j}^t$ is $A_i$'s prediction of $h_j^t$, and $h_{i,j,k}^t$ is $A_i$'s prediction of $h_{j,k}$. That is, $A_i$'s prediction of $A_j$'s prediction of $A_k$'s history.
    \item $Z$ is the set of all roles, and $\pi_{z}$ the policy of role z.
\end{itemize}

We also write $P_i$ to mean the probability that the $A_i$ assigns to an event, and $P_{i,j}$ to be $A_i$'s estimation of the the probability that $A_j$ would assign to an event. For example $P_{1,2}(A_1^{\text{r}})$ is $A_1$'s estimate of how much $A_2$ believes that $A_1$ is on the red team.

\subsection{The Model}

Given the assumptions in Section \ref{sec:assumptions}, we consider how a Bayesian agent $j$ would update their belief about agent $i$'s role given that they observed $A_i$ taking action $a$ in context $h_i$. We denote $A_i$'s true role with $*$.

\begin{equation}
    P_j(A_i^{*} |a, h_{i}) = 
        \frac
        {P_j(a | h_i, A_i^{*}) P_j(A_i^{*})}
        {P_j(a | h_i)}
\end{equation}

Because $A_j$ does not know $h_i$, they must estimate it as $h_{j,i}$. $A_i$ in turn does not know what $h_{j,i}$ is and so must estimate this too as $h_{i,j,i}$. Therefore $A_i$ can estimate $A_j$'s new belief about their role, if they were to take action $a$ as

\begin{equation}
    P_j(A_i^{*} |a, h_{i}) \approx
        \frac
        {P_{i,j}(a | h_{i, j,i}, A_i^*) P_{i,j}(A_i^*)}
        {P_{i,j}(a | h_{i, j,i})}
\end{equation}

Given that the probability of taking an action given history and role is, by definition, the known policies for each role, $A_i$ can estimate the Bayes factor with which $A_j$ would update their belief about $A_i$'s true role as

\begin{equation}
\label{eq:rho}
    \rho_{j,i} := \frac
        {\pi_{*}(a|h_{i,j,i})}
        {\sum_{z \in Z} \pi_z (a|h_{i,j,i})P_{i,j}(A_i^z)}
\end{equation}

As we wish to encourage deceptive agents to minimise this belief, we must reward agents for actions that generate ${\rho<1}$ and disincentivise actions that would generate ${\rho>1}$. Therefore, we provide agents with a $-\log(\rho)$ reward bonus, which fulfils this criterion, and whose interpretation is an estimate of the negative log of the Bayes factor in a Bayesian update for the other agent.

Implementing this bonus requires each agent $i$ to model both $h_{i,j,i}$ for each other player $j$, as well as model $P_{i,j}(A_i^z)$ for each $z \in Z$. The predicted history for each agent must then be run through each known policy to produce action distributions estimations for each role. 

\subsection{Direct Policy Prediction} 
\label{sec:dap}

The requirement that each agent must model all other other agents' complete histories can be removed by predicting the current policies for each agent directly. This avoids predicting irrelevant details about the other agents' observations, is more computationally efficient, and relaxes assumption (2) about needing to know the policies ahead of time.\footnote{Each player's current policy is needed as targets during training, so effectively this limitation is still in place. However, it is now a training detail rather than inherent to the model, and could potentially be removed by inferring policy from observed actions.} We do this by modelling $$\pi_{z}(h_{i,j,i})$$ directly for each role $z \in Z$ with $h_{i,j,i}$ being learned implicitly.  

This direct policy prediction method amounts to each agent predicting what agent thinks it would do in its current circumstance, if it was each of the known roles.

\subsection{Deception Bonus} 

Deceptive agents are given a bonus to their rewards based on their estimate of $\rho$ in equation \ref{eq:rho}. Similar to \citet{jaques2019social}, we filter out all but a 10\% random sample of non-visible agents as their contribution adds a lot of noise.\footnote{The argument that they did not observe the action and could therefore not update their belief is not valid here, as they could potentially see the consequences of the action in the future (a dead body for example). Our model does not handle this case and is left for future work.} We also exclude dead players from receiving or giving deception bonuses. The $\rho$ estimates for all remaining players are summed and added as an intrinsic reward as 

\begin{equation}
   r_i^t = r_{\text{ext}}^t + \alpha_i \times r_{\text{int}}^t 
\end{equation}

where $r_i^t$ is the reward at timestep $t$ for player $i$, and $\alpha_i$ is the magnitude of the deception bonus reward for $A_i$.

Therefore, agents are incentivised to take actions that would mislead other agents about their true role. This formulation has several advantages. First, agents are not incentivised to deceive players who already know their role, as in this case the Bayesian update would be very small. Second, agents are not incentivised to pretend to be another role when they can get away with it. If no other player is around, or if they believe that this action could not be interpreted negatively (as in Figure \ref{fig:theory_of_mind}), they will not be penalised. It is important to note, that in a complete information game all agents agree on all histories, and this situation would not occur.

\section{Training and Evaluation}

To evaluate the effect of BBM as an intrinsic motivation, we trained two groups of agents on the Rescue scenario. Each group consisted of four agents, trained with different random seeds. Agents in all groups trained with deception modules; however, the first group had $\alpha=0$ for all agents, while the second group set $\alpha=0.5$\footnote{$\alpha=0.5$ was chosen as initial tests showed this gave agents roughly one quarter of their reward from BBM. We kept the deception bonus reward small so as not to cause agents to loose sight of their primary objective, which was to win the game.} for red team players, and $\alpha=0$ for all other players. We trained the agents under the centralised learning, decentralised execution paradigm \cite{foerster2016learning} detailed further in Section \ref{sec:training}

\subsection{Agent Architecture}

\begin{figure*}
    \centering
    \includegraphics[width=0.90\textwidth]{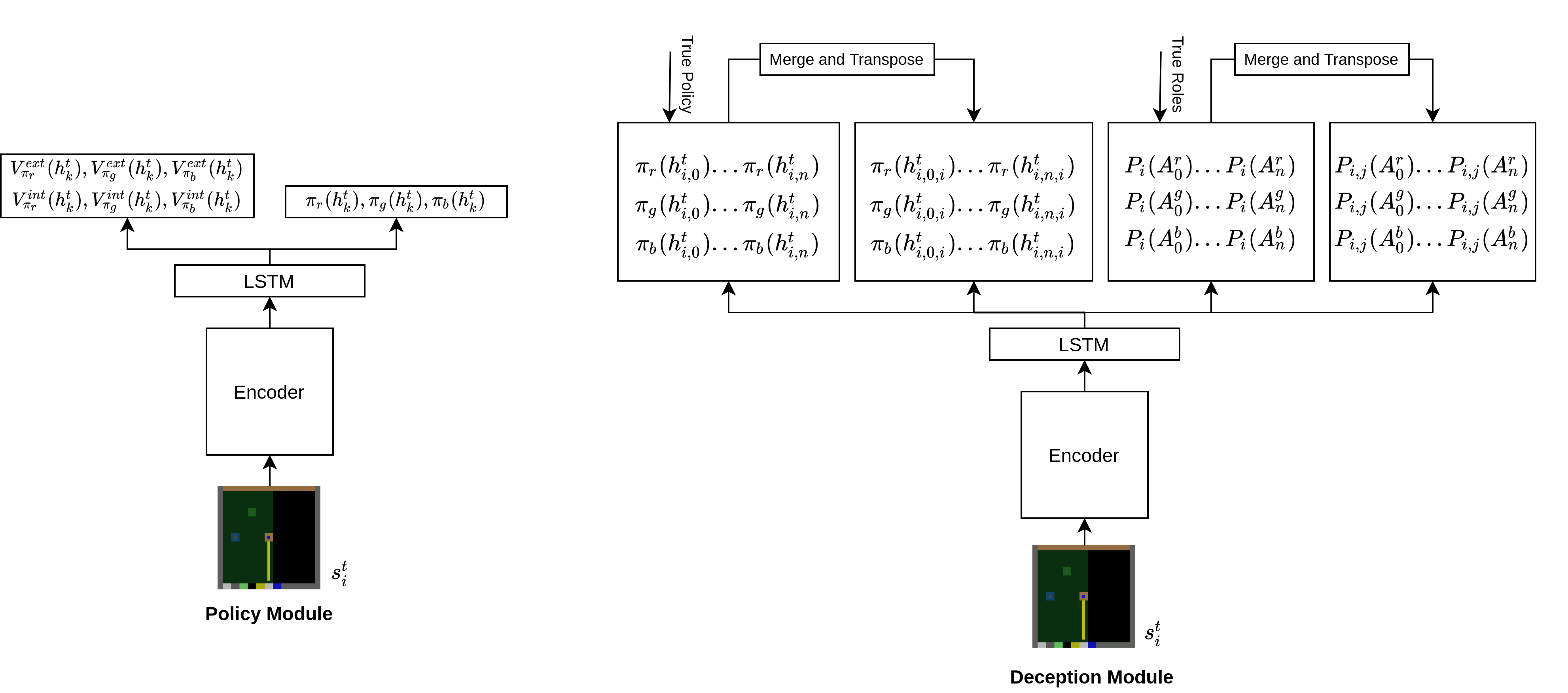}
    \caption{The architecture of the BBM agent. The policy module outputs policy and value estimations for each role, given the input history. The deception module outputs predictions and inverse predictions for all other players for both actions and roles and requires targets from other agents. The predictions for all players in a game are stacked and transposed, then used for targets so that each agent predicts the predictions of each other agent about themselves. Here $\pi_{r},\pi_{g},\pi_{b}$ are the current policy estimations for each player if they were red, green and blue. $V^{\text{ext}}$ and $V^{\text{int}}$ are the value estimations for the extrinsic and intrinsic returns respectively. $s^{t}_{i}$ is the local observation at timestep $t$ for the $i$th player, and $n$ is the number of players in the game. The output from the deception module is to generate deception bonus' only, and is not required once training is complete.}
    \label{fig:architecture}
\end{figure*}

Our BBM agent is composed of two distinct modules: the policy module and the deception module (see figure \ref{fig:architecture}). The policy module was trained using local observations only, while the deception module (discarded after training) required target information from other agents.

Both modules feature the same encoder architecture, which is loosely based on DQN \cite{mnih2013playing}, but modified to use 3x3 kernels to the match the 3x3 tile size and lower resolution of the RTG environment. To address partial observability a recurrent layer, in the form of Long Short Term Memory (LSTM) \cite{gers1999learning}, was added to the encoder, with residual connections from the convolution embedding to the output (see Appendix B for more detail on the encoder architecture.). 

\subsection{Training Details}
\label{sec:training}

The policy module takes an agent's local observation as input, encodes it via the encoder, and then outputs a policy and extrinsic/intrinsic value estimations for each of the three roles. During training, gradients are only backpropagated through the output head matching the role of the given agent.\footnote{This setup allowed us to record what an agent would have done if they were playing a different role.}

The policy module was trained using Proximal Policy Optimisation (PPO) \cite{schulman2017proximal} by running 4 epochs over batches of size 3,072 segments, each containing 16 sequential observations, using back propagation through time.

We also experimented with using separate models for each policy, but found little difference in the results, at the cost of longer training times (as did \citet{baker2019emergent}). A global value estimate based on the observations of all players was also considered, but we did not find significant differences in the estimates and so opted for local value estimation instead.

To make sure at least some games were winnable for each side we modified the game rules during training so that in half of the games, a player from a randomly selected team was eliminated before the game started. We found failing to do this sometimes caused one team to dominate the other, prohibiting progress.

The policy and deception modules were trained simultaneously on the same data. They share an identical encoder structure but used separate weights. Because of this, the deception module can be removed once training has finished, allowing for decentralized execution. The policy module produces the policy and value estimate based on local observations only. The deception module outputs predictions for player roles and player actions, as well as their inverse predictions (that is, what it predicts that other agents are predicting about itself). See Figure \ref{fig:architecture} for details. 

We trained each of the eight runs for 300-million interactions (which equates to 50-million game steps) over 5-days on four RTX2080 TIs using PyTorch 1.7.1 \cite{Paszke2019}. Checkpoints every 10-million interactions were saved, allowing for evaluation of the agents' performance at various points during training. 

Hyperparameters for the model, given in Table \ref{tab:hyperparameters}, were selected by running a grid search on the Red2 scenario. For more details refer to Appendix C.

\begin{table}[h]
    \small
    \sc
    \caption{Hyperparameters used during training.}
    \label{tab:hyperparameters}
    \centering
    \begin{tabular}{l l}
        \toprule
        \multicolumn{2}{c}{Policy Module} \\
        \midrule
        Mini-batch size & 128 segments of length 16\\
        LSTM units & 512\\
        PPO clipping $\epsilon$ & 0.2 \\
        Entropy coefficient & 0.01\\
        $\gamma$ & 0.99\\
        $\lambda$ & 0.95\\
        
        \midrule
        \multicolumn{2}{c}{Deception Module} \\
        \midrule
        Mini-batch size & 32 segments of length 8\\
        LSTM units & 1024\\
         
        \midrule
        \multicolumn{2}{c}{Shared} \\
        \midrule
        
        Parallel environments & 512 \\
        Learning rate & 2.5 $\times 10^{-4}$ \\
        Adam $\epsilon$ & 1 $\times 10^{-8}$\\
        Gradient Clipping & 5.0 \\
        \bottomrule
        
    \end{tabular}
    
\end{table}

\subsubsection{Losses}

Action predictions were trained by minimising the Kullback-Leibler (KL) \cite{kullback1951information} divergence between the predicted distribution and ground-truth targets taken from the agents during rollout generation. We trained the inverse action predictions in the same way, using the other agents' predictions during rollout generation as targets.

We trained the role predictions by minimising the negative likelihood loss, using players' true roles as targets. We trained inverse role predictions by minimising the KL divergence between the agent's prediction of other agents belief about its role, and their actual prediction about the agent's role. 

\subsubsection{Intrinsic Rewards}

Raw deception bonuses were calculated, clipped to the range (-20,20) and logged for each agent. Bonuses were then zeroed for all non-red players, normalised such that the intrinsic returns had unit variance, multiplied by $\phi(t)$ where 
$$\phi(t) = \min(1.0, t / 10 \times 10^6)$$
and $t$ is the timestep (in terms of interactions) during training then scaled by $\alpha_i$. This creates a warm-up period where agents can learn the game with without an auxiliary objective. Similar to \citet{burda2018exploration} our model outputs intrinsic and extrinsic value estimations separately, then combines them when calculating advantages.

\subsection{Evaluation}

Evaluating agents within a multi-agent scenario poses some unique challenges. Unlike games where agents compete against a static environment such as Atari, self-play means that as the agent gets stronger, so does its opponent. Thus, an agent's skill progress might not be visible from plots of the scores alone. 

To address this we evaluated each run by playing 16 games, every ten-epochs,\footnote{In this paper we define an epoch as one-million agent interactions with the environment.} against opponents from each of the eight runs (128 games in total). The green team was played using the same agent as the adversary. Performance across the four runs was averaged.

We also recorded raw deception bonuses, taken before normalisation or scaling, for all players, even those who did not receive them. This allowed for monitoring of deceptive behaviour on non-deceptively incentivised players. Accuracy of role predictions was taken from data generated during training.

\section{Effect of Deception on Agents' Behaviour}

In this section we present the effect of deception on the agents in terms of their ability to predict roles and their performance in the game, and discuss the tendency towards honest behaviour when no deception incentive is provided.

\subsection{Role Prediction}

\begin{figure}[H]
    \centering
    \includegraphics[width=0.75\textwidth]{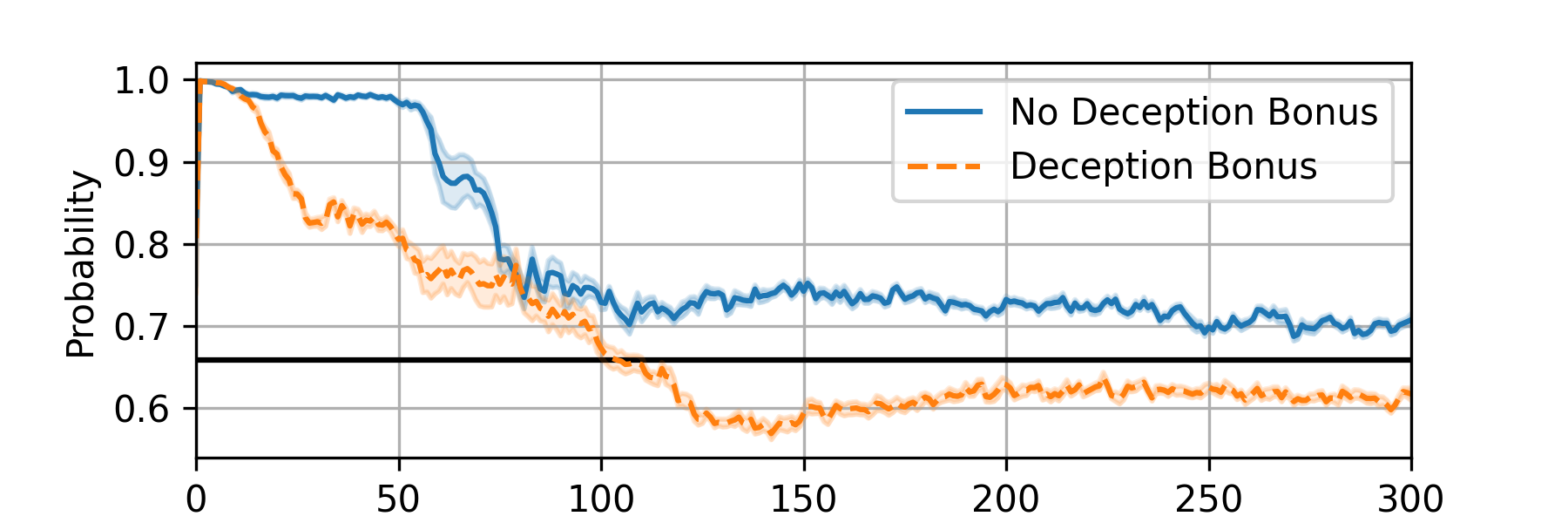}
    \caption{Probability assigned to true role the blue team. Values shown are the log-averages taken during training and represent the average probability a blue player assigned to each player's true role. As seen by the decrease in role prediction accuracy, the blue team struggles to identify players' roles when faced with deceptively motivated adversaries. The black line indicates a naive, count-based, estimation of player roles.}
    \label{fig:role_prediction}
\end{figure}

Identifying other agents' role is a difficult task for blue players in the Rescue scenario. At the end of training, blue assigns an average probability to players' true role of $0.617 \pm 0.08 (95\% \text{ CI})$ when trained against deceptive adversaries, as compared to $0.707 \pm 0.07 (95\% \text{ CI})$ when trained against non-deceptive adversaries (see Figure~\ref{fig:role_prediction}).\footnote{These percentages are derived from averages of the negative log likelihood loss taken during training, and are therefore \textit{geometric} averages.} Red's role predictions were very high at $0.995 \pm 0.001 (95\% \text{ CI})$, which is expected as they are able to see the roles of all visible players. A blue player who knows their role, and assigns remaining probability based on naive population counts would on average score 0.659. Therefore, the deceptive agents red caused the blue team to predict roles more poorly than if they had no in-game observations at all.

\subsection{Performance of Deceptive Agents}

\begin{figure*}[h]%
    \centering
    \subfloat[\centering Red vs Mixture ]{{\includegraphics[width=0.85\linewidth]{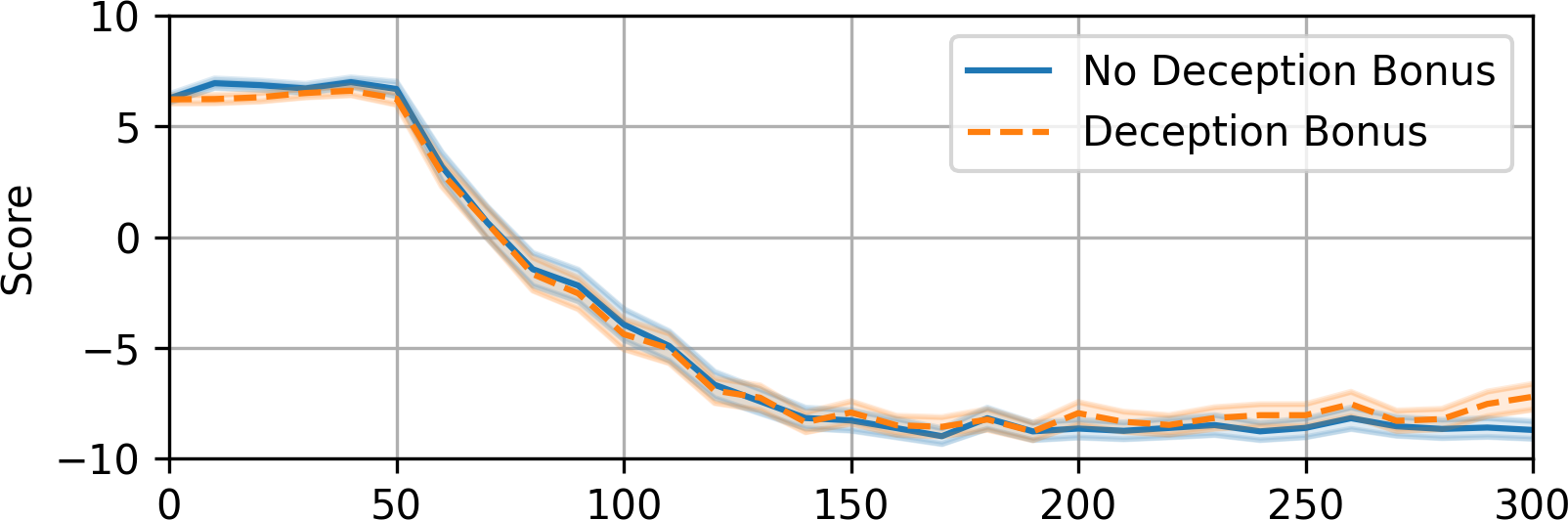} }}%
    \\
    \subfloat[\centering Blue vs Mixture]{{\includegraphics[width=0.85\linewidth]{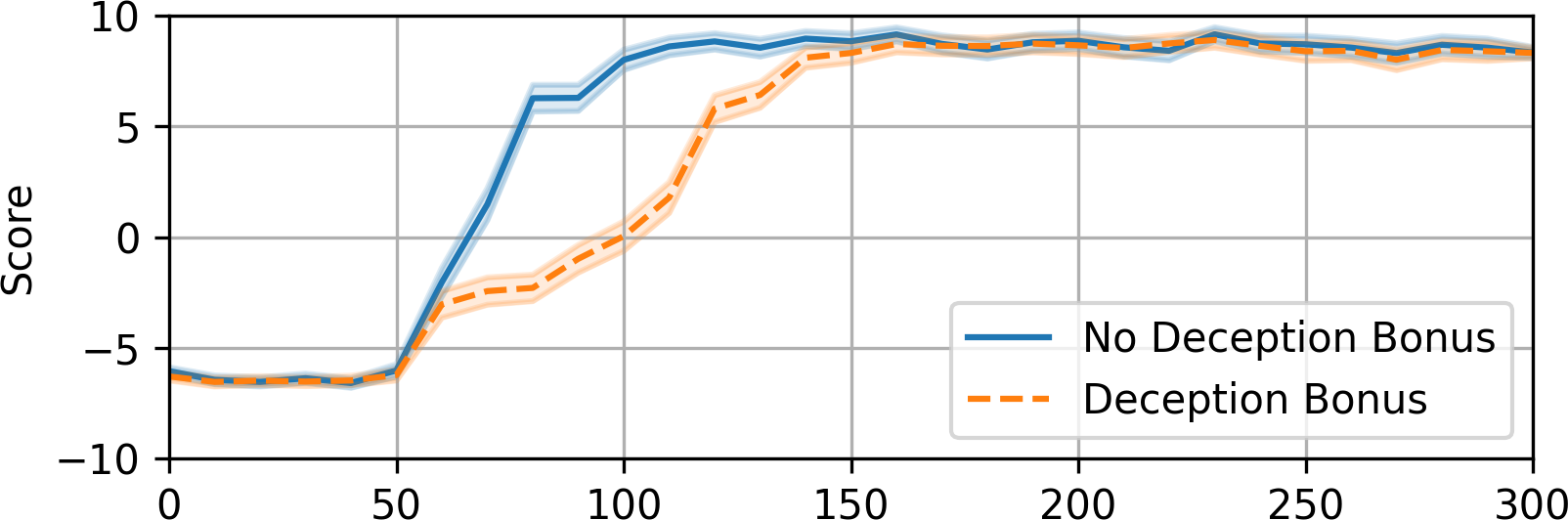} }}%
    \caption{The performance of the agents during training. Scores evaluated at 10-epoch intervals with opponents taken from a mixture of all eight runs. Shaded areas indicate 95\% confidence intervals.}%
    \label{fig:mixture}%
\end{figure*}

Despite the effectiveness on role deception, BBM has only a minor impact on the agents' score in our experiment. As seen in Figure \ref{fig:mixture} (a), red players trained with deception see an increase in performance at the end of training from $-8.71 \pm 0.040 (95\% \text{ CI})$ without deception to $-7.21 \pm 0.58 (95\% \text{ CI})$ with deception. This result, while statistically significant, is only seen at the end of training, and does not represent a large difference in performance. It does, however, suggest further experimentation into the effect of a larger deception bonus. When facing deceptive agents blue is slower to learn a winning strategy, but eventually converges to the same outcome (see Figure \ref{fig:mixture} (b)).

\subsection{Deceptive Tendencies in Unincentivised Agents}

We look now at agents' behaviour when not given explicit deception incentives. Blue learns to act honestly when faced with either deceptive or honestly adversaries. This is expected in the Rescue scenario as blue must identify teammates in order to coordinate with them. An unexpected result is that green, who initially acts honestly, learns an increasingly deceptive strategy, despite no explicit incentive to do so. We suspect that as red develops a strategy that involves pretending to be green, green must respond by pretending not to be green to survive blue's hostility.

\begin{figure*}[t]%
    \centering
    \subfloat[\centering With Deception ]{{\includegraphics[width=0.85\linewidth]{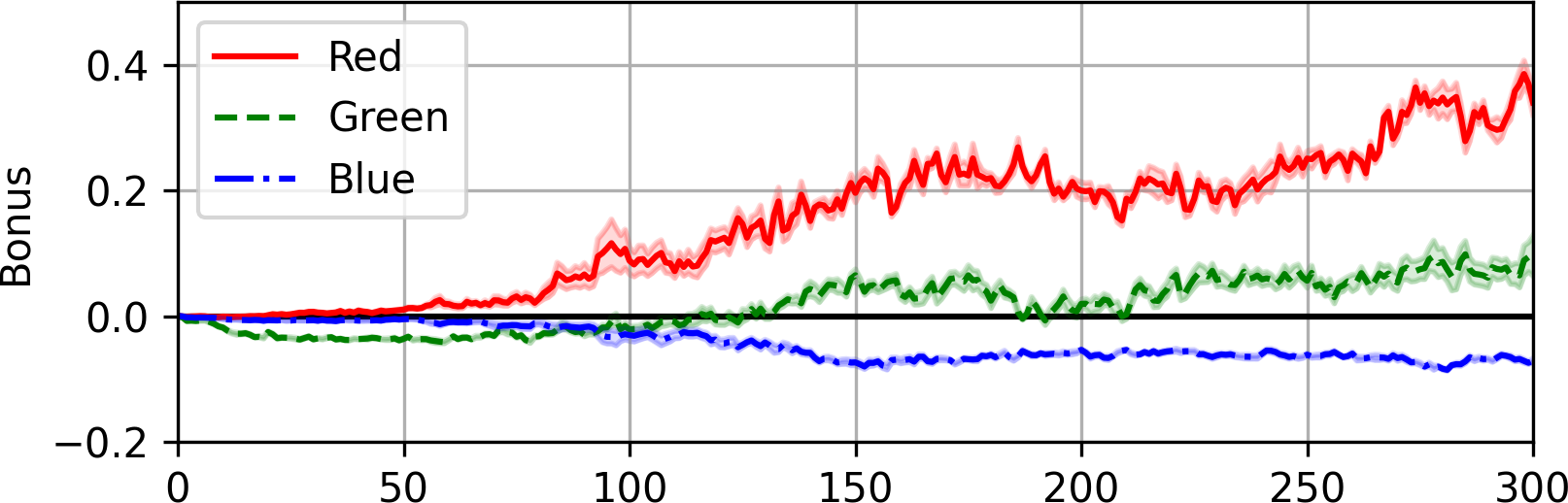} }}%
    \\
    \subfloat[\centering Without Deception]{{\includegraphics[width=0.85\linewidth]{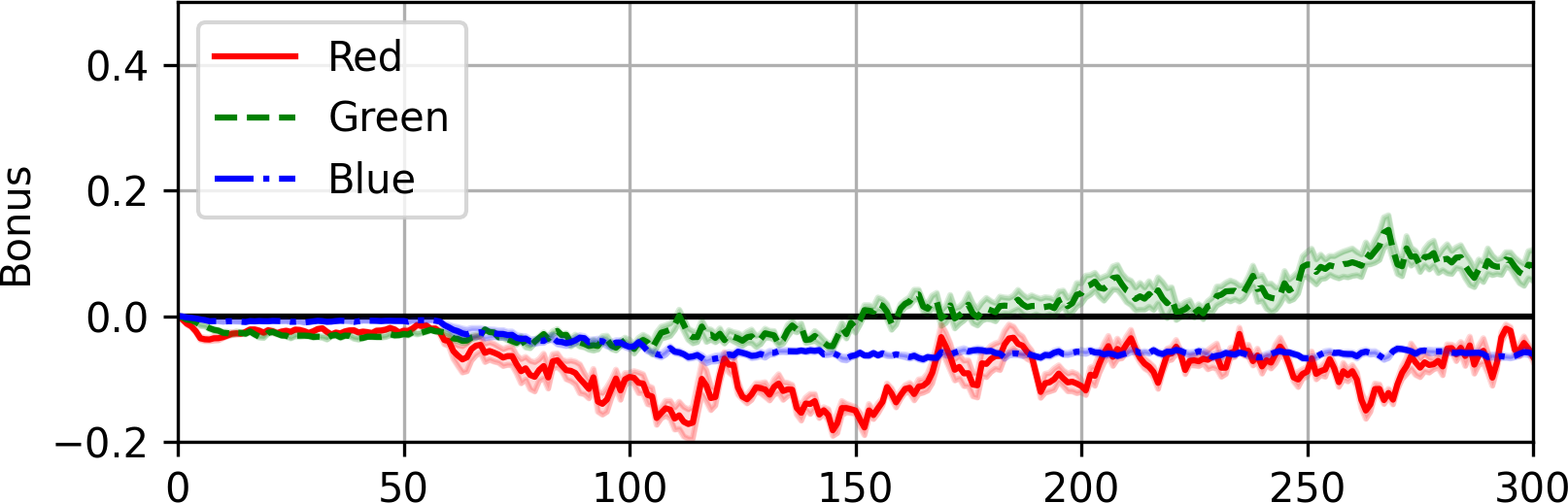} }}%
    \caption{Raw deception bonuses for each team.  These are the unscaled, unnormalised intrinsic rewards that each player would have received if they were to receive a bonus. Values are averaged across all four runs, with the shaded area representing a 95\% confidence interval. The black line indicates behaviour which would neither reveal, nor deceive other players about one's role.}%
    \label{fig:raw_deception}%
\end{figure*}

The red players, when incentivised to do so, learn to obtain a high degree of deception bonus (see Figure \ref{fig:raw_deception} (a)). When no explicit incentive is provided red players do not naturally learn a deceptive strategy. This suggests that either deception is of less value for the red team in the Rescue scenario than expected, or that a deceptive, but effective, strategy is difficult to learn for red.

\section{Conclusions and Future Work}
Our work presents a new, open-source environment, RTG for development and testing of deception in a complex hidden role game, as well as a novel a model of deception that makes use of second-degree ToM to manipulate other agents belief about it's own role.  Our empirical results show BBM to be very effective at manipulating other agents belief about their role while at the same time improving their performance in the game. We also found that deceptive behaviour, when not incentivised, is not learned naturally by the competitive red or blue teams in the Rescue scenario, but surprisingly is learned by the neutral green team. 

These results suggest several questions for future research into the effect of deception in a MARL setting. Can the relatively modest performance improvement in performance be extended by increasing the magnitude of the deception bonus, or would this cause agents to act deceptively at the cost of their primary objective? What would the impact of an honesty bonus be on the blue team in the Rescue scenario, would this encourage better cooperation? How would incorporating social aspects, namely explicit communication, impact deceptive behaviour? And how could our model, which assumes only one deceptive team, be extended to account for two or more deceptive teams.

\section*{Acknowledgements}

This initiative was funded by the Department of Defence and the Office of National Intelligence under the AI for Decision Making Program, delivered in partnership with the NSW Defence Innovation Network.

\newpage
%
%
%
\bibliographystyle{splncs04}
\bibliography{references.bib}

\appendix


\section{Scenario Configuration Options}\label{app:config}

We designed Rescue the General (RTG) to be highly customisable through configuration options. We used only a subset of the configuration settings in the supplied scenarios, with the full set outlined in Table \ref{tab:config}. We also designed a voting system where players could initiate votes to remove players. However, we found that agents never developed an effective strategy for this system, and so did not use it.

\section{Encoder Architecture}\label{app:encoder}

\begin{figure}[h]
    \centering
    \includegraphics[width=0.25\textwidth]{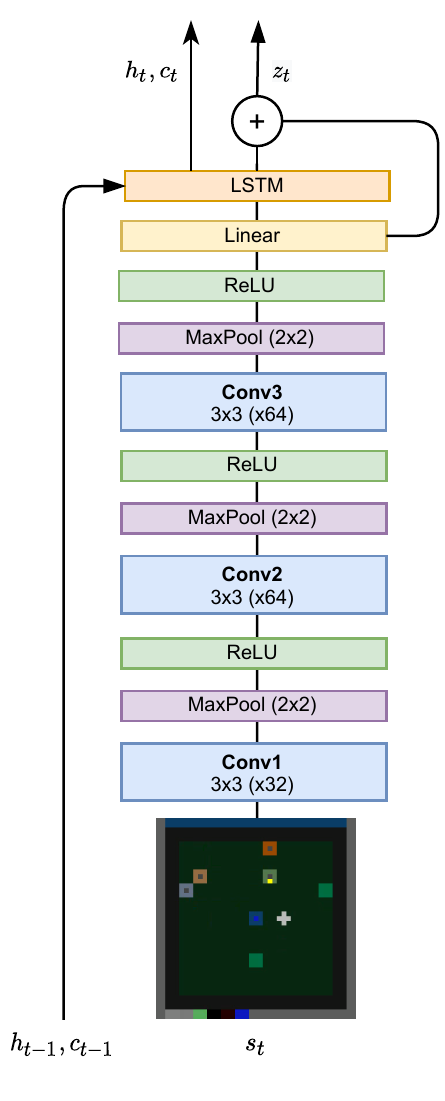}
    \caption{Encoder Architecture. Input $s_t$ is fed into a convolutional neural network, a linear layer, and finally, an LSTM layer.}
    \label{fig:encoder}
\end{figure}

In our model, both the policy module and deception module share the same encoder architecture, but with separate parameters. The encoder design is a modification of the network often used in Deep Q-Networks (DQN), but with smaller 3x3 kernels to match our game's tile-size. The encoder takes a single local observation scaled to $[0..1]$, $s_t$,  as input, as well as the LSTM cell states $(h_{t-1}, c_{t-1})$ from the previous timestep. The input is fed through the convolutional layers, flattened and projected into $n$ dimensions, where $n$ are the number LSTM units, then fed through the LSTM layer. See Figure \ref{fig:encoder} for details. We found that adding a residual connections skipping the LSTM layer improved training performance, and did not negatively affect the agents' ability to remember past events.\footnote{We verified this by observing that red players were able to retain information about the roles of previously seen players} The policy module used 512 LSTM units, whereas the deception module used 1024.

\section{Hyperparameter Search}\label{app:search}

We performed a hyperparameter search for the policy module using a randomized grid-search. Our search used 128 runs over ten epochs on the Red2 scenario.\footnote{We did not use the Rescue scenario as it is more difficult to evaluate performance, and runs would require significantly more training time.} We used the average score taken from the last 100 episodes during training, where $$\text{score} := R \times 0.99^{t}$$ with $R$ being the red teams score at the end of the episode and $t$ being the episode length. Discounting score in this way prefers hyperparameter settings that can more quickly find and kill the general and distinguishes between runs that consistently score the optimal ten points.

The \textit{out features} setting controls the number of units in the final linear layer. When \textit{residual} mode was active \textit{out features} were set to the same value as \textit{LSTM units}. The \textit{off} mode disabled the LSTM layer, \textit{on} passed through the LSTM layer as per normal, \textit{concatenate} concatenated the LSTM output with the linear layer, and \textit{residual} added residual connections.

Hyperparameter values were selected during the search by sampling uniformly-random from the values in Table \ref{tab:search_pm}. We then plotted each hyperparameter by selecting the best five runs for each value setting and graphing their min, mean and max. 

\begin{table}[h]
    \centering
    \small
    \caption{Hyperparameters tested for policy module. Mini-batch sizes refer to the number of observations, not segments. N-steps is the number of steps used when generating rollouts.}
    \begin{tabular}{l p{4.5cm} l}
        \toprule
        Parameter & Values & Selected \\
        \midrule
        N-steps & 16, 32, 64, 128 & 16 \\
        Parallel Environments & 128, 256, 512, 1024 & 512\\
        Learning Rate & $10^{-4}$, $2.5 \cdot 10^{-4}$, $10^{-3}$ & $2.5 \cdot 10^{-4}$ \\
        Adam $\epsilon$ & $10^{-5}$, $10^{-8}$ & $10^{-8}$ \\
        Gradient Clipping & off, 0.5, 5.0 & 5.0  \\
        Mini-Batch-Size & 64, 128, 256, 512, 1024, 2048 & 2048  \\
        LSTM Mode & off, on, concatenate, residual & residual \\
        LSTM Units & 64, 128, 256, 512, 1024 & 512 \\
        Out Features & 64, 128, 256, 512, 1024 & 512 \\
        Discount $\gamma$ & 0.95, 0.99, 0.995 & 0.99 \\
        Entropy Coefficient & 0.003, 0.01, 0.03 & 0.01 \\
         \bottomrule
    \end{tabular}
    \label{tab:search_pm}
\end{table}

The purpose of our hyperparameter search was not to find optimal settings but to find reliable settings that would allow us to assess the difference between deceptively incentivised agents, and the control agents. Therefore, in cases where hyperparameters had similar performance, we preferred values used in prior work and those with better computational efficiency.

A separate hyperparameter search was used to find hyperparameters for the deception module, using 64 runs over 25 epochs on R2G2 with hidden roles. This search was performed using an older version of the algorithm where agents predicted other agents observations rather than action distributions and used the negative log of the mean-squared prediction error as the score. We found these settings to work well on the updated algorithm and did not modify them. Details are given in Table \ref{tab:search_dm}. 

We enabled window slicing (\textit{max window size}) on the deception module. During deception module training, shorter windows were extracted from the longer rollout segments. This allowed the deception module to use a shorter back-propagation-through time length than the policy module, which we found to be effective. We also found that the deception module benefited from much smaller mini-batch sizes (256) than the policy module (2048). Due to differences in scale between the role prediction loss and the action/observation prediction loss, we multiplied action/observation prediction losses by \textit{loss scale}.

\begin{table}[h]
    \small
    \centering
    \caption{Hyperparameters tested for deception module. Settings omitted used values found from the policy module search.}
    
    \begin{tabular}{l p{3.5cm} l}
        \toprule
        Parameter & Values & Selected \\
        \midrule
        Learning Rate & $10^{-4}$, $2.5 \cdot 10^{-4}$, $10^{-3}$ & $2.5 \cdot 10^{-4}$ \\
        Mini-Batch-Size & 128, 256, 512, 1024 & 256  \\
        LSTM Units & 128, 256, 512, 1024 & 1024 \\
        Loss Scale & 0.1, 1, 10 & 0.1 \\
        Max Window Size & 1, 2, 4, 8, 16, 32 & 8 \\
        
         \bottomrule
    \end{tabular}
    \label{tab:search_dm}
\end{table}

\begin{table*}[b]
    \centering
    \caption{Configuration settings for Rescue the General}
    \begin{tabular}{l p{7.5cm} c}
        \toprule
         Parameter & Description & Default \\
         \midrule
         map\_width & Width of map in tiles. & 32 \\
         map\_height & Height of map in tiles. & 32 \\ 
         
         n\_trees & Number of trees on map. & 10 \\
         reward\_per\_tree & Points for green for harvesting a tree. & 1\\
         max\_view\_distance & Distance used for observational space, unused tiles are blanked out. & 6\\%
         team\_view\_distance & View distance for each team & (6,5,5) \\
         team\_shoot\_damage & Damage per shot for each team. & (10, 10, 10) \\
         team\_general\_view\_distance & Distance a player must be from the general to see them for the first time. & (3, 5, 5) \\
         team\_shoot\_range & Distance each team can shoot. & (5, 5, 5) \\
         team\_counts & Number of players on each team. & (1, 1, 4) \\
         team\_shoot\_timeout & Number of turns between shooting & (10, 10, 10) \\
         
         enable\_voting & Enables the voting system. & False \\
         voting\_button & Creates a voting button near start location. & False \\
         
         auto\_shooting & Removes shooting in cardinal directions and replaces it with a single action that automatically targets the closest player. & False \\
         
         zero\_sum & If enabled any points scored by one team will be counted as negative points for all other teams. &  False   \\
        
         timeout & Maximum game length before a timeout occurs. & 500 \\
         
         general\_initial\_health & General's starting health. & 1 \\
         player\_initial\_health & Players' starting health. & 10 \\
         
         battle\_royale & Removes general from the game. Instead, teams win by eliminating all other players. & False \\  
            
         help\_distance & How close another player must be to help the first player move the general. & 2 \\
        
         starting\_locations &
         \textit{random} - starts players at random locations throughout the map.\newline \textit{together} - places players in a group around a randomly selected starting location. & together \\
            
        local\_team\_colors & If enabled, team colours are included on players local observation.
 & True \\
        
        initial\_random\_kills & Enables random killing of players at the start of the game. & 0.5 \\ 
        
        blue\_general\_indicator & \textit{direction} - blue team is given direction to general.\newline \textit{distance} - blue team is given distance to general. & direction\\
        
        players\_to\_move\_general & Number of players required to move the general. & 2 \\
        
        timeout\_penalty & Score penalty for each team if a timeout occurs. & (5,0,-5) \\
        
        points\_for\_kill & Matrix $A$ where $A_{i,j}$ indicates points player from team $i$ receives for killing a player on team $j$. & 
        $\mathbf{0}$
        \\

        hidden\_roles & \textit{default} - red can see roles, but blue and green cannot.\newline \textit{all} - all players can see roles. \newline \textit{none} - no players can see roles. & default\\ 
        
        reveal\_team\_on\_death & Enables display of team colors once a player dies & False\\

         \bottomrule
    \end{tabular}
    \label{tab:config}
\end{table*}

\end{document}